\documentclass[sigconf]{acmart}

\usepackage{balance}

\usepackage{booktabs} 
\usepackage{graphicx}
\usepackage{subfigure}
\usepackage{paralist}
\usepackage{amsmath}
\usepackage{amssymb}
\usepackage{algorithm}
\usepackage{algorithmic}
\usepackage{array}
\usepackage{url}
\usepackage{multirow}
\usepackage{natbib}
\usepackage{bm}
\usepackage{enumitem}
\usepackage{amsthm}
\newcolumntype{L}{>{\centering\arraybackslash}m{6.5cm}}
\newtheorem{myDef}{Definition}

\def\BibTeX{{\rm B\kern-.05em{\sc i\kern-.025em b}\kern-.08emT\kern-.1667em\lower.7ex\hbox{E}\kern-.125emX}}

\copyrightyear{2019}
\acmYear{2019}
\setcopyright{acmlicensed}
\acmConference[KDD '19] {The 25th ACM SIGKDD Conference on Knowledge Discovery and Data Mining}{August 4--8, 2019}{Anchorage, AK, USA}
\acmBooktitle{The 25th ACM SIGKDD Conference on Knowledge Discovery and Data Mining (KDD '19), August 4--8, 2019, Anchorage, AK, USA}
\acmPrice{15.00}
\acmDOI{10.1145/3292500.3330922}
\acmISBN{978-1-4503-6201-6/19/08}

\begin{document}


%
\title{Exploiting Cognitive Structure for Adaptive Learning}



\author { 
    Qi Liu$^1$, Shiwei Tong$^1$, Chuanren Liu$^2$,  Hongke Zhao$^3$, Enhong Chen$^{1,*}$, 
}
\author{
    Haiping Ma$^{4,5}$, Shijin Wang$^{4,5}$
}

\affiliation{
	\institution{$^1$Anhui Province Key Laboratory of Big Data Analysis and Application, School of Computer Science and Technology \& School of Data Science, University of Science and Technology of China,\\ \{qiliuql, cheneh\}@ustc.edu.cn, tongsw@mail.ustc.edu.cn}
}

\affiliation{
	\institution{$^2$Business Analytics and Statistics, University of Tennessee, chuanren@xminer.org}
}

\affiliation{
	\institution{$^3$The College of Management and Economics, Tianjin University, hongke@tju.edu.cn}
}


\affiliation{
	\institution{$^4$iFLYTEK Research, iFLYTEK CO., LTD., $^5$State Key Laboratory of Cognitive Intelligence, \{hpma,sjwang3\}@iflytek.com}
}

\begin{abstract}
Adaptive learning, also known as adaptive teaching, relies on learning path recommendation, which sequentially recommends personalized learning items (e.g., lectures, exercises) to satisfy the unique needs of each learner.
Although it is well known that modeling the cognitive structure including \textit{knowledge level} of learners and \textit{knowledge structure} (e.g., the prerequisite relations) of learning items is important for learning path recommendation, existing methods for adaptive learning often separately focus on either knowledge levels of learners or knowledge structure of learning items.
To fully exploit the multifaceted cognitive structure for learning path recommendation, 
we propose a Cognitive Structure Enhanced framework for Adaptive Learning, named \textbf{CSEAL}.
By viewing path recommendation as a Markov Decision Process and applying an actor-critic algorithm, CSEAL can sequentially identify the right learning items to different learners.
Specifically, we first utilize a recurrent neural network to trace the evolving knowledge levels of learners at each learning step.
Then, we design a navigation algorithm on the knowledge structure to ensure the logicality of learning paths, which reduces the search space in the decision process.
Finally, the actor-critic algorithm is used to determine what to learn next and whose parameters are dynamically updated along the learning path. Extensive experiments on real-world data demonstrate the effectiveness and robustness of CSEAL.
\end{abstract}

%
%


%

\keywords{
    Adaptive Learning; Knowledge Graph; Reinforcement Learning
    \let\thefootnote\relax\footnotetext{$^*$Corresponding Author.}
}

\maketitle

\section{Introduction}

\begin{figure*} [ht]
    \footnotesize
	\centering
	\includegraphics[width=0.83\textwidth]{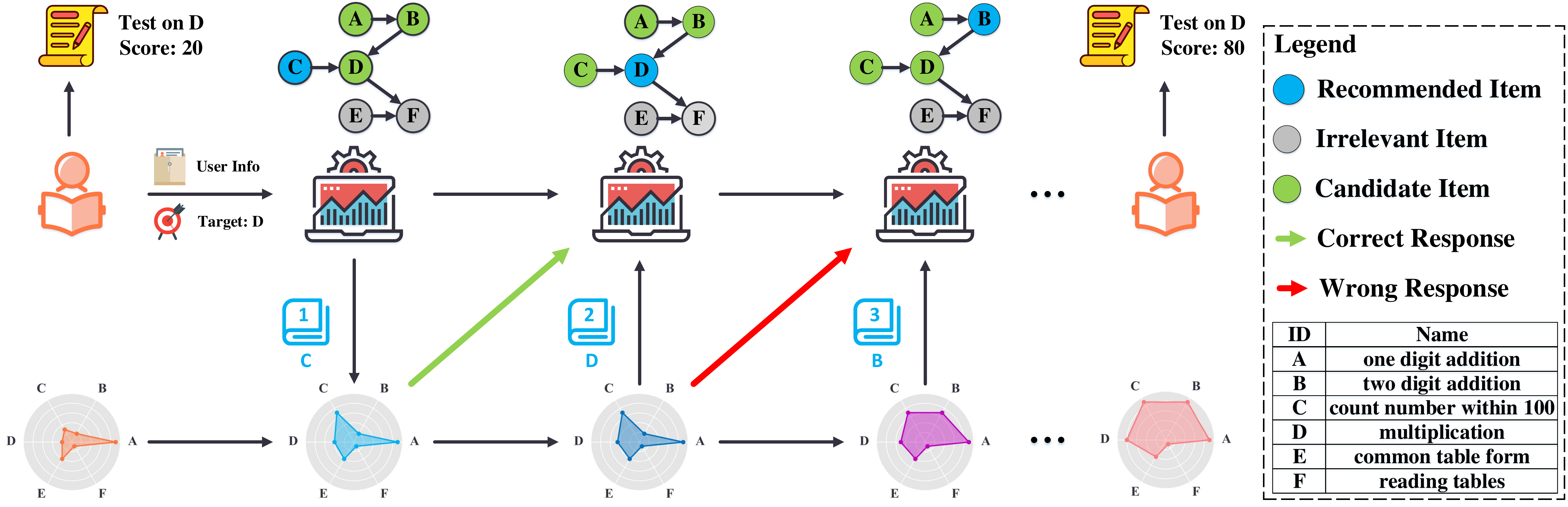}\vspace{-0.5cm}
	\caption{
	    Illustration of Adaptive Learning. C $\rightarrow$ D $\rightarrow$ B $\rightarrow$ ... is a learning path which promotes the learner's mastery on D from 20 pt to 80 pt. User Info contains the basic information of learners such as grade, historical learning records. Target includes one or more learning items that the learner wants to master. Radar graphs in the bottom show the evolving knowledge level during learning which can not be directly observed, and the top directed acyclic graphs represent the knowledge structure.
	}
	\label{fig:AdaptiveLearning}\vspace{-0.5cm}
\end{figure*}

\textit{Learning is the ladder of the progress of mankind}, by which people can acquire knowlege and skills.
Different from traditional learning (e.g. courses in classrooms) that presents the same material to all learners, adaptive learning aims at providing personalized learning items and paths tailored to individual learners \citep{carbonell1970ai}.
As shown in Figure \ref{fig:AdaptiveLearning}, adaptive learning recommends a learning path C $\rightarrow$ D $\rightarrow$ B $\rightarrow$ $\cdots$ for the learner who wants to learn \textit{multiplication} by considering his or her current level of knowledge and the prerequisite relation of learning items (e.g., \textit{two digit addition} is a prerequisite of \textit{multiplication}). Usually, examinations, e.g., two tests in Figure \ref{fig:AdaptiveLearning} on D, are used to retrieve the learning effects.
The personalized learning path helps the learner understand the new learning items efficiently \citep{tang2018reinforcement,liu2018fuzzy}.
Recently, adaptive learning has become a crucial component for many applications such as on-line education systems (e.g., \textit{KhanAcademy.org}, \textit{junyiacademy.org}).

Research on education has shown that the cognitive structure has great impacts on adaptive learning \cite{ piaget1976piaget, piaget1970genetic, pinar1995understanding}.
The cognitive structure describes the qualitative development of knowledge and contains two parts: knowledge level of learners and knowledge structure (e.g., the prerequisite relations) of learning items.
The knowledge level reflects the masteries on learning items which keeps evolving and can not be observed directly (e.g. genetic epistemology ~\cite{piaget1970genetic}), meanwhile the knowledge structure captures the cognitive relations among the learning items.
However, existing methods for adaptive learning only utilize either knowledge level ~\cite{zhou2018personalized,tang2018reinforcement} or knowledge structure ~\cite{zhu2018multi, yu2007ontology}.
Although these methods have made a great success in adaptive learning, there are some limitations of them.
To be specific, 
methods based on knowledge level without the knowledge structure may fail to resolve learning items dependency, e.g., prerequisite.  The methods based on knowledge structure which ignore learners' knowledge level can not reflect the learning abilities of different learners, so that they can not precisely determine the customized learning tempo~\cite{chen2018recommendation}. Therefore, the recommended learning paths to each learner may be less suitable and inefficient.
Thus, how to systematically exploit cognitive structure including both knowledge level and knowledge structure for adaptive learning is still a challenging problem.

We summarize three challenges along this line.
First, the knowledge level of learner cannot be observed directly and keeps evolving.
As shown in the radar graphs of Figure ~\ref{fig:AdaptiveLearning}, the learner's masteries of each learning items, i.e. the knowledge level, are continuously changing during learning but can not be observed directly.
The knowledge level of learners influences the learning effectiveness of learning items. For example, with the poor mastery of \textit{two digit addition}, it is difficult to learn \textit{multiplication}.
Thus, it is necessary to model the implicit evolving knowledge level.
Second, the knowledge structure of learning items should be incorporated to determine logical learning paths \citep{liu2011mining}.
As shown in the directed acyclic graphs of Figure ~\ref{fig:AdaptiveLearning}, a learner who wants to learn the senior item D (\textit{multiplication}) should firstly learn the prerequisites B (\textit{two digit addition}) and C (\textit{count number within 100}).
Therefore, the learning path should be in accordance with the logicality determined by the knowledge structure of learning items.
Third, a good learning path recommendation should maximize the overall gain along the whole learning path instead of only focusing on the gain in one step.
As shown in Figure~\ref{fig:AdaptiveLearning}, the learning effectiveness is the promotion in examinations rather than simply the correction of one item \citep{chen2018recommendation}.

To address the challenges, we propose a general Cognitive Structure Enhanced framework for Adaptive Learning (\textbf{CSEAL}).
We model the sequential learning path recommendation as a Markov Decision Process (MDP).
We apply reinforcement learning to gradually optimize the recommendation strategy based on cognitive structure of adaptive learning.
Specifically, the Knowledge Tracing model based on Long Short-Term Memory (LSTM) network is first applied to retrieve the evolving knowledge level.
Second, to prevent learning path from violating the sequential logicality being recommended, the Cognitive Navigation algorithm is designed to select a certain number of learning items, i.e., candidates, based on knowledge structure, which can also reduce the large searching space. Finally, the Actor-Critic Recommender will determine what to learn next, whose parameters are updated to improve the effectiveness of the whole recommended learning path rather than that of only one item.
Extensive experiments show that CSEAL not only significantly outperforms several baselines, but also provides interpretable insights in the learning path recommendations.

\vspace{-0.3cm}
\section{Related Work}

Generally, the related work of this study can be grouped into the following three categories.

\textbf{Learning Path Recommendation.}
The simplest way to generate learning paths is to introduce those methods aiming to solve sequence recommendation problem, e.g., collaborative filtering methods (e.g., KNN ~\cite{cover1967nearest}, MPR ~\cite{yu2018multiple}) and deep learning methods (e.g. GRU4Rec ~\cite{DBLP:journals/corr/HidasiKBT15}). 
For example, Zhou et al. ~\cite{zhou2018personalized} introduced Recurrent Neural Network (RNN) to predict the expectation of the whole path for learner groups. 
Some researches proposed to enhance the recommendation strategy by explicitly using cognitive structure. One branch is to model the evolution of knowledge level. Chen et al. ~\cite{chen2018recommendation} and Tang et al. ~\cite{tang2018reinforcement} used the transition matrix in MDP to model the evolution of knowledge level and used reinforcement learning algorithm to evaluate the impact of learning items on knowledge level. Meanwhile, works of the other branch focused on employing the knowledge structure to make a recommendation, for example, Zhu et al. ~\cite{zhu2018multi} made several path generation rules on knowledge structure by expertise and Yu et al. ~\cite{yu2007ontology} put up a method using semantic inference on ontology to generate learning paths. 
Previous methods only consider either the importance of knowledge level or knowledge structure without the combination of these two parts. To our best, few of existing works has well established the cognitive structure to make learning path recommendation.

\textbf{Cognitive Structure.}
Cognitive structure contains two parts: knowledge level of learners and knowledge structure of items. Two kinds of techniques can be applied to describe these two components, i.e., knowledge tracing for knowledge level and knowledge graph for knowledge structure. Knowledge tracing models learners' knowledge level over time so that how learners will perform on future interactions can be accurately predicted~\cite{corbett1994knowledge,chen2017tracking,su2018exercise}. 
Deep Knowledge Tracing (DKT) ~\cite{piech2015deep} used RNN to model such states in a high-dimensional and continuous representation. However, the mentioned-above knowledge tracing approaches are hard to fully reveal the relations among the learning items.

Knowledge Graph, with entities (e.g. learning items) as nodes, relations (e.g. prerequisite) as edges, stores a large amount of information containing domain knowledge by graph structure ~\cite{nickel2016review}. Meanwhile, the education knowledge graph is able to represent the multi-dimension relationships~\cite{DBLP:conf/icdm/Chen0ZP18, liu2018finding,huang2017question}. Though abundant of knowledge can be thus embedded in Knowledge Graph, the personalized evolution of knowledge level's characteristic, especially the complexities and the dynamic specialty with infinite time horizons, makes it difficult to be described in the graph structure.

\textbf{Reinforcement Learning.} 
Deep reinforcement learning, as one of state-of-the-art techniques~\cite{DBLP:journals/corr/abs-1708-05866}, has shown superior abilities in many fields ~\cite{yin2018transcribing}. The main idea is to learn and refine model parameters according to task-specific reward signals. For example, Tang et al. ~\cite{tang2018improving} introduced reinforcement learning to train an efficient dialogue agent on existing transcripts from clinical trials, which improves mild cognitive impairment prediction; Wang et al. ~\cite{wang2018supervised} utilized the actor-critic algorithm for treatment recommendation, helping to handle complex relations among multiple medications, diseases and individual characteristics. However, due to three key challenges, the traditional reinforcement learning is difficult to be applied in learning path recommendation: (1) how to represent state; (2) how to avoid the recommendation violating the sequence logicality during exploring; (3) how to reduce the large searching space of learning item paths.

\vspace{-0.2cm}
\section{PRELIMINARIES}
This section discusses the definition of the terminologies and the formulation of learning path recommendation. 
\vspace{-0.3cm}
\subsection{Terminologies}

\subsubsection{Learning Session} 
\begin{figure} [t]
	\footnotesize
	\begin{center}
		\includegraphics[width=0.35\textwidth]{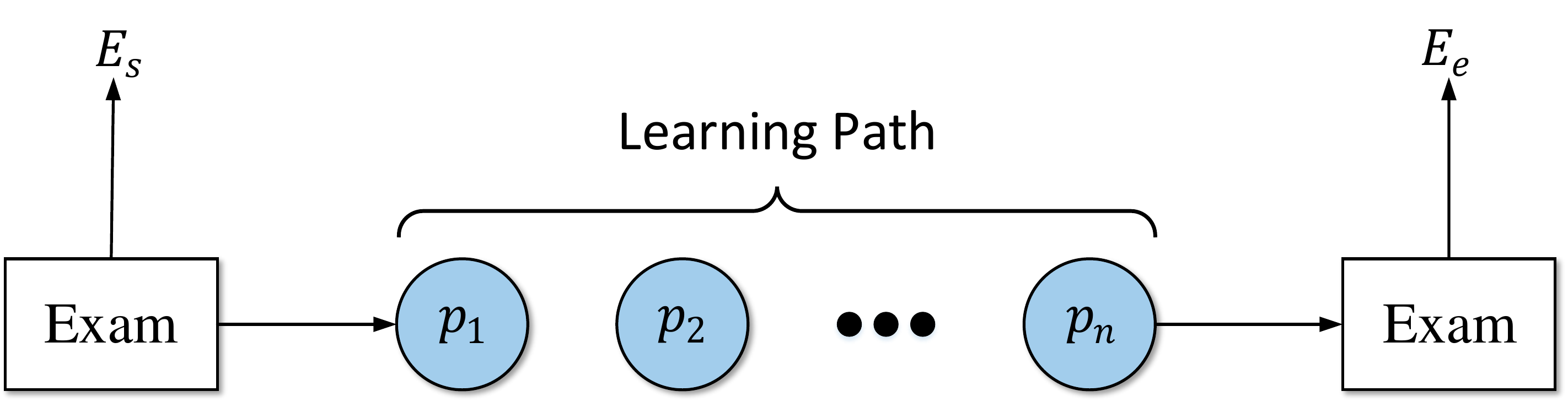}
	\end{center}\vspace{-0.45cm}
	\caption{Illustration of a Learning Session.}
	\label{fig:Session}\vspace{-0.1cm}
\end{figure}

Learning is a long procedure composed of many learning sessions, and each learning session has individual learning targets which can be set by tutors or learners. As shown in Figure ~\ref{fig:Session}, each learning session includes two main components: learning path and examinations. Learning path consisting of many learning items is a learning track of a learner. Examinations are used to retrieve the learning effectiveness of the learning path, i.e., the promotion on the learning target. Some typical learning sessions are homework, chapters and semesters, which differ in gratitude. Without loss of generality, the effectiveness of a learning session $E_{\mathcal{P}}$ can be calculated by the following equation:
\begin{small}\vspace{-0.1cm}
    \begin{equation} \label{eq:EP}
    E_\mathcal{P}= \frac{E_e - E_s}{E_{sup} - E_s},
    \end{equation}
\end{small}where $E_s$ is the score of the beginning examination in a session, $E_e$ is the score of the end, and $E_{sup}$ is the full score of the examination. For example, let $E_s=80, E_e=90, E_{sup}=100$ and then $E_\mathcal{P}=0.5$, or let $E_s=0.2, E_e=0.6, E_{sup}=1.0$, then $E_\mathcal{P}=0.5$.

\vspace{-0.1cm}
\subsubsection{Prerequisite Graph} An educational knowledge graph has some special properties such as prerequisite and similarity, which can represent the knowledge structure. As learners often start from basic items before accessing to those senior ones which are more complicated and hard ~\cite{DBLP:conf/icdm/Chen0ZP18}, experts hence summarize a relation of learning items, named prerequisite. A prerequisite graph is a subgraph of knowledge graph, which indicates the hierarchical structure existing among learning items. As shown in directed acyclic graphs of Figure ~\ref{fig:AdaptiveLearning}, the nodes of the graph represent learning items while the arrows from one to another mean that the former is a prerequisite for the latter, e.g., \textit{two digit addition} is a prerequisite for \textit{multiplication}.

\vspace{-0.2cm}
\subsection{Problem Formulation}
As mentioned above, learning is composed of many learning sessions. In each session, a learner will try to master a specific learning target $\mathcal{T}=\{t_0, t_1, ...\}$, which contains one or more learning items. The learning items are the nodes on a prerequisite graph $G$ where the edge represents the prerequisite relation, i.e., $(i, j)$. The tuple $(i, j)$ indicates that the learning item $i$ is a prerequisite of $j$. For a learner who is going to begin a new session, the historical learning records generated in previous learning sessions are denoted as $\mathcal{H}=\{h_0, h_1, ..., h_m\}$. Each record $h_i=\{k, score\}$ contains a learning item $k$ and the corresponding performance $score$. Without loss of generality, we assume $score$ as a discrete number 0 or 1, where 1 indicates the corresponding item is correctly answered , while 0 stands in the opposite. Our goal is to recommend an optimized learning path $\mathcal{P}=\{p_0,p_1,...p_N\}$ containing $N$ items to the learner sequentially, by which can the learner achieve a greater promotion. Specifically, at step $i$, a learning item $p_i$ is recommended and the interaction learning record $\mathcal{F}_i=\{p_i, score_i\}$ can be observed. At the end of the learning session, we can calculate the learning effectiveness $E_{\mathcal{P}}$. After all, the problem is defined as:

\vspace{-0.2cm}
\begin{myDef} (Learning Path Recommendation Problem) Given historical learning records $\mathcal{H}$, a certain learning target $\mathcal{T}$ of a learner and a prerequisite graph $G$, our task is to recommend a $N$-length learning path $\mathcal{P}$ step by step that can maximize the effectiveness $E_{\mathcal{P}}$ of the whole learning path. During recommendation, we can observe a new interaction learning record $\mathcal{F}_i$ of each recommended learning item $p_i$ instantly.
\end{myDef}
\vspace{-0.4cm}

\begin{figure*} [t]
    \footnotesize
	\centering
	\includegraphics[width=0.8\textwidth]{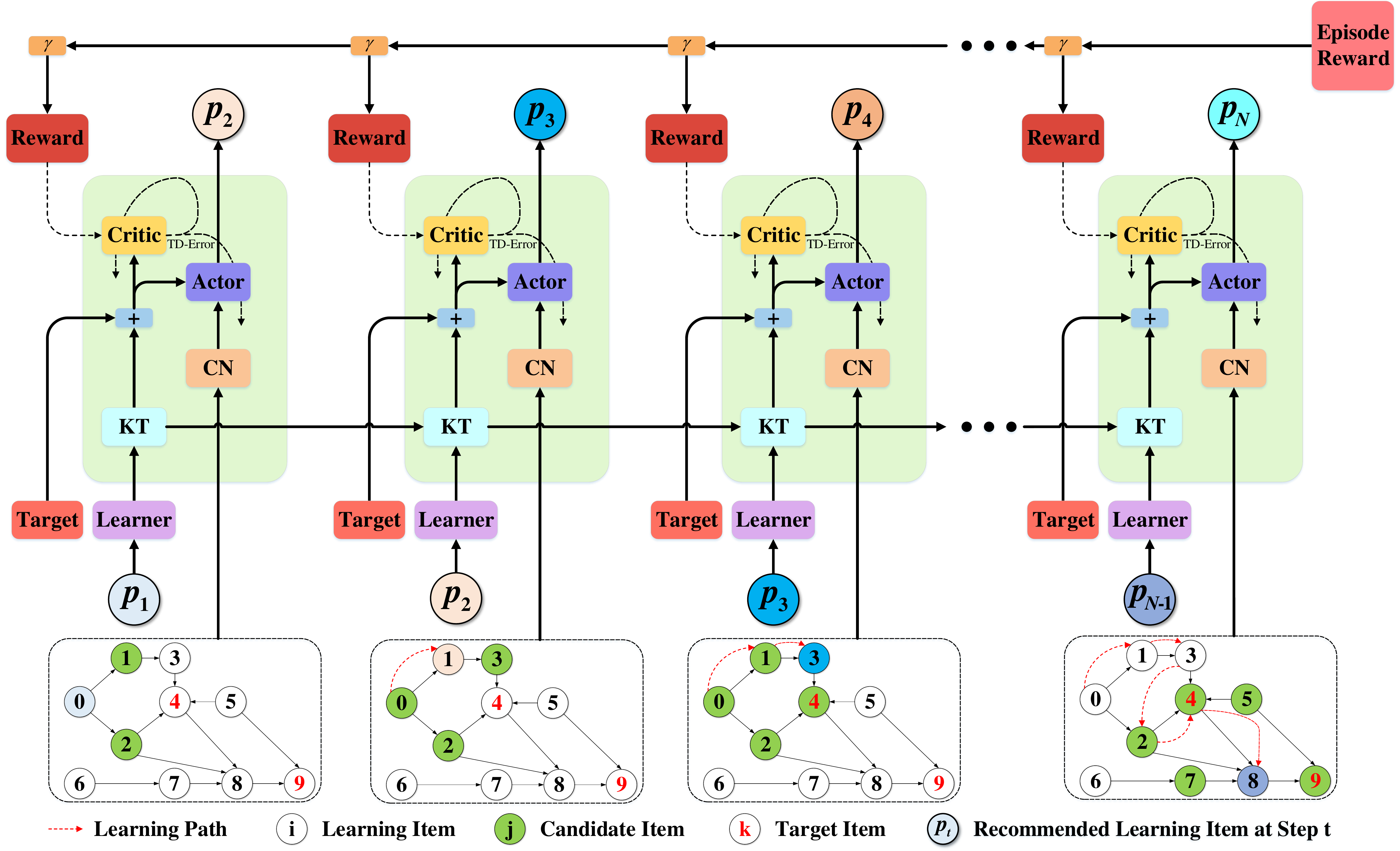}\vspace{-0.45cm}
	\caption{The overview of our framework.}
	\label{fig:Framework}\vspace{-0.5cm}
\end{figure*}

\section{CSEAL}
This section begins with a brief overview of our framework with the definition of the MDP. The details of CSEAL is then introduced. 

\vspace{-0.2cm}
\subsection{Overview} \label{sec:Framework}
The target is to learn a policy to recommend tailored learning paths based on the cognitive structure. We model such sequential path generation as a decision making problem and treat it as a MDP ~\cite{bahdanau2016actor}. The state, action and reward of the MDP are defined as follows:

\textbf{State}. Generating the probability of learning items at each step is based on the learning target and previous learning records. The state at step $i$ is represented as the combination of the learning target $\mathcal{T}$ and current knowledge level $\mathcal{S}_i$, which are combined and denoted as $state_i$. Specifically, we use one-hot encoding to signify the learning target as $\mathcal{T}={\{0, 1\}}^M$:
\begin{small}
    \vspace{-0.1cm}
    \begin{equation}
    \mathcal{T}^j = \left\{
    \begin{array}{cl}
    1   & \text{if $j$ in the learning target} \\
    0   & \text{otherwise}, \\
    \end{array} \right.
    \end{equation}
\end{small}where $M$ is the number of nodes in the prerequisite graph $G$. However, the current knowledge level can not be observed directly, thus we need to find a way to retrieve it from the previous learning records $\mathcal{L}_{i-1}$ including historical learning records $H$ and previous interaction learning records $\mathcal{F}_{0,...,i-1}$.

\textbf{Action}. Taking action $a_i$ refers to generating the recommended learning item $p_i$ at step $i$. With the probability of each item as output, the CSEAL can be viewed as a stochastic policy that generates actions by sampling from the distribution $\pi(a|state_i;\theta)=P(a|\mathcal{H}, \mathcal{F}_{0,...,i-1},\mathcal{T};\theta)$, where $\theta$ is the set of model parameters.

\textbf{Reward}. After taking the action, a reward signal $r$ is received. We determine the reward $r_i$ at step $i$ keeps to be 0 during learning session. Once the learning session is completed, the reward is calculated by Equation ~(\ref{eq:EP}). Our goal is to maximize the sum of the discounted rewards from each step $i$. That is, the return:
\begin{small}
    \begin{eqnarray}\label{eq:reward}
    R_i = \sum_{i}^{N}\gamma^ir_i,
    \end{eqnarray}
\end{small}where $\gamma$ is the discount factor, which is usually set to 0.99.

Generally, our CSEAL model contains three submodules, i.e., Knowledge Tracing (KT), Cognitive Navigation (CN), Actor-Critic Recommender (ACR). As shown in Figure ~\ref{fig:Framework}, KT retrieves the knowledge level $\mathcal{S}$ from previous learning records at each step and CN selects several candidates based on the prerequisite graph $G$. With the learning target and knowledge level composing the state, ACR determines what to learn next by maximizing the overall gain along the whole learning path. At the end of the episode, i.e., learning session, an episode reward will be passed to CSEAL and used in reinforcement learning stage.

\vspace{-0.2cm}
\subsection{Kowledge Tracing} \label{sec:KT}
\begin{figure} [t]
	\scriptsize
	\begin{center}
		\includegraphics[width=0.4\textwidth]{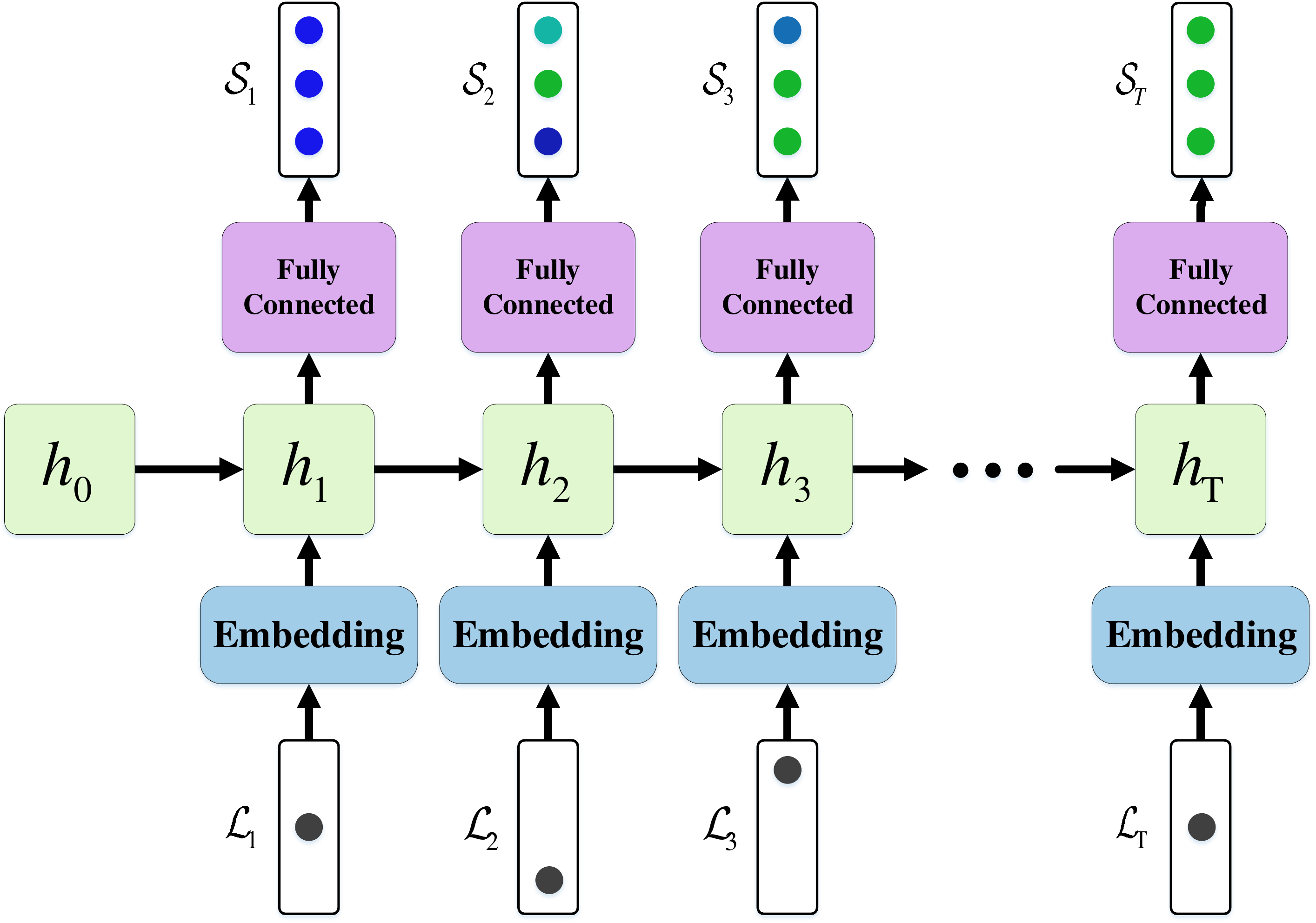}
	\end{center}\vspace{-0.45cm}
	\caption{Illustration of an embedding DKT.}
	\label{fig:DKT}\vspace{-0.1cm}
\end{figure}

Knowledge level of learners does greatly influence the strategy of recommending learning path. Also it is a key component of state in MDP, which should be well established. However, learner's knowledge level is unobservable and evolving. To this end, Knowledge Tracing is applied to retrieve the implicit knowledge level $\mathcal{S}_i$ from previous learning records $\mathcal{L}_{i-1}=\mathcal{H} \oplus \mathcal{F}_{0,...,i-1}$. We follow the structure proposed by the original work of Deep Knowledge Tracing (DKT) ~\cite{piech2015deep} and extend an embedding layer which reduces the large feature space. Figure ~\ref{fig:DKT} is a cartoon illustration of the architecture of this module. Without the loss of generality, we assume $score_t$ in each record of previous learning records $\mathcal{L}_t=\{p_t, score_t\}$ is either 0 or 1. We use one-hot representation to stand for $\mathcal{L}_t$ as described in DKT. 

Firstly, an embedding operation is utilized to convert the one-hot representation of the record into a low-dimensional one. Formally, for a record $\mathcal{L}_t$, the converted vector $u_t$ is expressed as:
\begin{small}
    \begin{equation}
    x_t=\mathcal{L}_tW_u.
    \end{equation}
\end{small}Here, $W_u \in \mathbb{R}^{2 \cdot M \times d}$ indicates the parameters of the embedding layer and $x_t \in \mathbb{R}^{d}$, where $d$ is the output dimension.

After obtaining the feature representation of a learning record, KT aims at tracing the knowledge level along the learning records. An LSTM architecture as described in DKT is then used to map the input sequence of vectors $x_1,x_2,...,x_N$ to the output sequence of hidden knowledge level vectors $o_1, o_2, ..., o_N$. The hidden state $h_t$ at the $t$-th input step is updated as following formulas:
\begin{small}
    \begin{eqnarray}\label{eq: LSTM}
    && i_t = \sigma(\mathbf{W}_{xi}x_{t} + \mathbf{W}_{hi}h_{t-1} + \mathbf{b_i}), \nonumber  \\
    && f_t = \sigma(\mathbf{W}_{xf}x_{t} + \mathbf{W}_{hf}h_{t-1} + \mathbf{b_f}), \nonumber  \\
    && o_t = \sigma(\mathbf{W}_{xo}x_{t} + \mathbf{W}_{ho}h_{t-1} + \mathbf{b_o}), \nonumber  \\
    && c_t = f_t c_{t-1} + i_t tanh(\mathbf{W}_{xc}x_{t} + \mathbf{W}_{hc}h_{t-1} + \mathbf{b}_c), \nonumber  \\
    && h_t = o_t tanh(c_{t}).
    \end{eqnarray}
\end{small}Meanwhile, by using a fully connected layer to retrieve the knowledge level from  $o_1, o_2, ..., o_N$, vector representation of knowledge level $\mathcal{S}_t \in \mathbb{R}^M$ can be expressed as:
\begin{small}
    \begin{eqnarray}\label{eq:vf}
    \mathcal{S}_t = \sigma(\mathbf{W}_{o\mathbb{F}}o_t + b_\mathbb{F}),
    \end{eqnarray}
\end{small}where $i_\bullet$, $f_\bullet$, $c_\bullet$, $o_\bullet$ are the input gate, forget gate, memory cell, output gate of LSTM respectively. $\mathbf{W_\bullet}$ and $\mathbf{b_\bullet}$ are learned weight matrices and biases.

\vspace{-0.3cm}
\subsection{Cognitive Navigation} \label{sec:PM}
The learning items have some inherent semantic structure characteristics, i.e., knowledge structure, which should be maintained to make sure the paths are logical sequences. Intuitively, dependency resolution methods \cite{kahn1962topological,zhao2019voice} can be applied. However, such methods require clear conditions for the resolution, which is hard to be satisfied in learning path recommendation due to the complexity of combination of knowledge level and knowledge structure. Thus, we focus on quickly selecting the potential candidates. These potential candidates would not only prevent recommended learning items from violating the logicality of the path, especially during exploring (e.g., avoid exploring the effect of recommending \textit{calculus} to junior students) but also reduce the large searching space.

With the prerequisite graph under $p_3$ in Figure ~\ref{fig:Framework} as an example, when a learner finishes the learning item 3, he or she is arranged to review the prerequisites (e.g., item 0, item 1, item 2) or continue to preview the the following item 4. We call the just learned item as ``central focus" (i.e., item 3) and those items which can be then possibly selected to learn as ``candidates" (e.g., item 0, item 1, item 2 and item 4). Simply, these candidates can be chosen from the neighbors of "central focus" which are potential to resolve the dependency of learning targets, more precisely, reach the learning targets. The algorithm runs following the steps shown in Algorithm ~\ref{alg:PG}.

\begin{small}
    \vspace{-0.2cm}
    \begin{algorithm}
	\caption{Cognitive Navigation.}
	\label{alg:PG}
	\begin{algorithmic}[1]
		\REQUIRE central focus $\mathcal{C}$, a prerequisite graph $G$, learning target $\mathcal{T}$
		\ENSURE $\forall d \in D$ has a path to $\mathcal{T}$ in $G$ and is one of the $k$-hop neighbors of $\mathcal{C}$
		\STATE Initialize candidates $\mathcal{D} = \varnothing, Q=\varnothing$; 
		\STATE add $\mathcal{C}$ to $\mathcal{D}$;
		\STATE add successors within $k-1$ hop of $\mathcal{C}$ to $\mathcal{D}$;
		\STATE add predecessors within $k-1$ hop of $\mathcal{C}$ to $Q$;
		\WHILE{$Q \neq \varnothing$}
		\STATE $q \leftarrow Q$.pop();
		\STATE add $q$ to $D$;
		\STATE add neighbors of $q$ to $D$;
		\ENDWHILE
		\FOR{$d$ in $D$} 
		\IF{$d$ can not reach $T$}
		\STATE del $d$ from $D$;
		\ENDIF 
		\ENDFOR
		\RETURN $\mathcal{D}$
	\end{algorithmic}
    \end{algorithm}\vspace{-0.4cm}
\end{small}

By setting $k$ as 2, the bottom part of Figure ~\ref{fig:Framework} illustrates how central focus and candidates change with path generating where the green ones are the candidates of next step. The central focus $\mathcal{C}$ at the first step can be assigned as the last item of $\mathcal{H}$, which can also be specified manually or chosen from any nodes without predecessors. The Cognitive Navigation is well capable to maintain the logicality of the path during generation. We set $k=2$ in following analysis. 

\vspace{-0.3cm}
\subsection{Actor-Critic Recommender} \label{sec:AS}
Although we achieve the two parts' information of the cognitive structure, there still exists a problem: which candidate item should be chosen based on the current knowledge level of the learner. To solve this problem, we use a policy network as the actor to generate actions among $\mathcal{D}$ given by Section~\ref{sec:PM} through sampling from the distribution $\pi(a|state_i;\theta)$ and a value network as the critic to evaluate the state. The value network $\mathcal{V}(\cdot ; \theta_\mathcal{V})$ is used to estimate the expected return from each state. It is a feed-forward network whose input is $state_i=\mathcal{S}_i \oplus \mathcal{T}$, where $\mathcal{S}_i$ is the current knowledge level given by Section~\ref{sec:KT} and $\mathcal{T}$ is the learning target. The estimated expected return $\mathcal{V}_i$ at step $i$ is computed by:
\begin{small}
    \begin{eqnarray}\label{eq:value_net}
    v_i = \mathcal{V}(state_i; \theta_\mathcal{V}) = \mathcal{V}(\mathcal{S}_i \oplus \mathcal{T}; \theta_\mathcal{V}).
    \end{eqnarray}    
\end{small}

With a stochastic policy together with a value network, we apply the actor-critic algorithm ~\cite{konda2000actor, bahdanau2016actor} to our sequential generation problem, with the policy network trained using policy gradient at each step $i$ as:
\begin{small}
    \begin{equation}
    \nabla_\theta = log\ \pi(a|state_i;\theta)(R_i-v_i),
    \end{equation}
\end{small}and the value network trained by optimizing the distance between the estimated value and actual return:
\begin{small}
    \begin{equation}
    \mathcal{L}oss_{value} = {\left \| v_i - R_i \right\|}_2^2.
    \end{equation}    
\end{small}

A too fast convergence of the value network may result in the slow convergence or even no-convergence of policy network. We therefore raise a policy enhanced loss item to address this issue, and the loss function is thus formulated as:
\begin{small}
    \begin{equation}
    \begin{aligned}
    \mathcal{L}oss =
    & {\left \| \mathcal{V}(state_i;\theta_\mathcal{V}) - R_i \right\|}_2^2 + \alpha\cdot -log\ \pi(a|state_i;\theta)(R_i-v_i) \\
    & + \beta\cdot -log\ \pi(a|state_i;\theta)R_i,
    \end{aligned}
    \end{equation} 
\end{small}where $\alpha$ and $\beta$ are the hyper-parameters.

%

\vspace{-0.2cm}
\section{Experiments}
In this section, we first introduce the dataset. Then, we train agents of reinforcement learning models and evaluate recommended learning paths in two kinds of environments. At last, the performance of our framework is compared with several baselines.

\vspace{-0.2cm}
\subsection{Dataset Description} \label{sec:DS}
\begin{figure} [t]
	\scriptsize
	\begin{center}
		\includegraphics[width=0.45\textwidth]{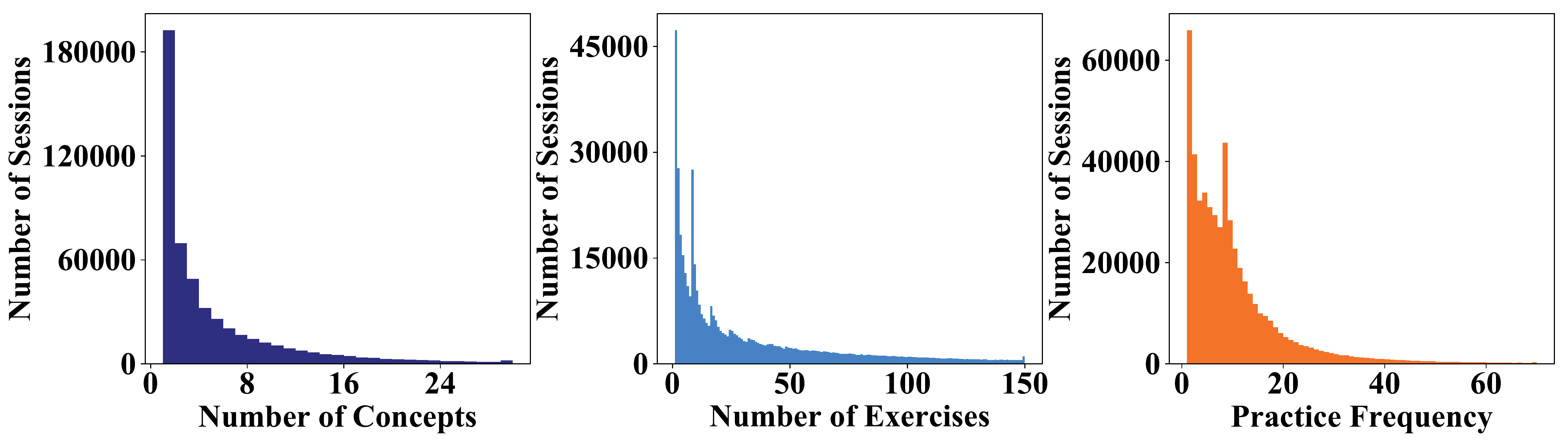}
	\end{center}\vspace{-0.45cm}
	\caption{Distributions of Sessions.}
	\label{fig:session_data}\vspace{-0.1cm}
\end{figure}

The experiment dataset is from \textit{junyiacademy.org} and collected by Chang et al. ~\cite{chang2015modeling}. The dataset includes a knowledge graph and more than 39 million learners' logs. Each record in the learners' log contains the information of a learner for one exercise, i.e., user id, concept name, session id, correction and time stamp. The exercise can be mapped to a node in the knowledge graph based on the concept name. And each exercise and concept is one-to-one correspondence \footnote{In some works the exercise and concept may be one-to-many correspondence ~\cite{chen2017tracking,piech2015deep}, while others has the same correspondence realtionship as ours ~\cite{DBLP:conf/icdm/Chen0ZP18,chen2018recommendation}.}. Those records with the same session id represent that the practiced data contributed by the same learner in one session. Grouped by session id and sorted by time stamp, session learning records can be extracted (e.g., \{(\textit{representing\_numbers}, correct), (\textit{division\_4}, wrong), (\textit{conditional\_statements\_2}, wrong), (\textit{conditional\_statements\_2}, wrong)\}). We further extract a prerequisite graph from the knowledge graph. The prerequisite graph contains several edges, e.g., (\textit{one\_digit\_addition}, \textit{two\_digit\_addition}) stands for the linkage between the node \textit{one\_digit\_addition} and the node \textit{two\_digit\_addition} where the former is the prerequisite of the latter. We delete some loop in order to keep the graph to be a Directed Acyclic Graph (DAG), which indeed is the knowledge structure. The preprocessed dataset are detailed in Table ~\ref{tab:data} and Figure ~\ref{fig:session_data}. Observed from the data, three key notes should be emphasized: 
(1) The length of more than 75\% sessions are longer than 6;
(2) more than 75\% sessions contain one more concepts;
(3) the median of practice frequency on one concept in one session is 8.
It conclusively infers that a concept in one session may be practiced repeatedly and some relevant concepts are learned simultaneously. In other words, during a learning path of the session, multiple concepts and the related ones contribute to the final learning result.

\vspace{-0.3cm}
\subsection{System Simulators} \label{sec:SS}
A key problem is that existing realistic data only contains static information, i.e., several exercise sequences and whether a certain exercise is answered correctly. This information can not be directly employed to analyze whether an exercise not included in a certain exercise sequence can be answered correctly. Thus the realistic data can not be directly used as the environment to evaluate the learning paths (e.g. calculating the promotion) or to train the agents (e.g. CSEAL and other baselines) in reinforcement learning models. It is therefore important to construct a simulator as the environment which can model qualitative development of knowledge and the performance on a certain learning item. More precisely, at each episode, the environment can simulate a learner whose knowledge level can be changed by the recommended path. The knowledge levels of the learner will be measured by the proposed environment at the beginning and the end of the learning session. Hence, $E_s$, $E_e$ and $E_{sup}$ are obtained by the simulators. The promotion of the level (i.e, $E_\mathcal{P}$) therefore can be calculated. We refer the ideas constructing simulators in state-of-art methods of not only education ~\cite{chen2018recommendation, tang2018reinforcement} but also other areas, like transportation ~\cite{DBLP:conf/kdd/WeiZYL18, DBLP:conf/kdd/LiZY18}, e-commerce ~\cite{DBLP:conf/kdd/HuDZ0X18}. Following these works, we raise two ways to build the simulators.

\begin{table}[t]
	\small
	\centering
	\caption{The statistics of the dataset.}\vspace{-0.4cm}
	\label{tab:data}
	\begin{tabular}{c|c}
		\hline
		Statistics & Value \\
		\hline
		number of learners & 247,548 \\
		number of sessions & 525,062 \\
		number of learner logs & 39,462,202 \\
		number of exercises in learner logs correctly answered & 21,460,360 \\
		median of exercises in one session & 21 \\
		median of knowledge concepts in one session & 3 \\
		median of practice frequency on a concept in one session & 8 \\
		\hline
		number of nodes in graph & 835 \\
		number of links in graph & 978 \\
		\hline
	\end{tabular}\vspace{-0.3cm}
\end{table}


\textbf{Knowledge Structure based Simulator(KSS).} According to previous works ~\cite{chen2018recommendation,tang2018reinforcement,piech2015deep}, we design a simulator where the pattern of qualitative knowledge development fits perfectly for knowledge structure. More specifically, in this simulator, the masteries on prerequisites do affect the successors, e.g., poor mastery of \textit{two digit addition} impairs the learning effectiveness of \textit{multiplication}. To scale the relation of mastery and learner performance, we use a widely applied method in education, Item Response Theory (IRT) ~\cite{lord1952theory,lord2012applications}. The 3-parameter logistic model of IRT is formulated as:
\begin{equation}
P(\theta) = c + \frac{1-c}{1+e^{-Da(\theta-b)}},
\end{equation}where $D=1.7$ is a constant; $c$ is a pseudo-guessing parameter; $\theta$ is the mastery of the learner on a certain learning item; $a$ is the item discrimination; $b$ is the item difficulty; and $P(\theta)$ is the probability of correctly answering the corresponding exercise which is used in computing $score_i$ for each $p_i$ and calculating the reward. The relevant parameters and evolving rules of $\theta$ are designed by experts in education which makes KSS a rule-based expert system.
%
%
%
%
%

\textbf{Knowledge Evolution based Simulator(KES).} We propose a data-driven method to construct a simulator which can better approximate learners' knowledge level and be abbreviated as knowledge evolution. We train a DKT model based on the existing data. The input of DKT is a record data, and the output, $\mathcal{S}_i$, is the current knowledge level of the learner. Then whether the next exercise can be answered correctly is determined by this knowledge level. 
More precisely, the probability of an exercise answered correctly is treated as its mastery value. For example, the probability of an exercise belonged to $i$ being answered correctly is $\mathcal{S}_i$. The probability will be used to
compute $score_i$ for each $p_i$ and calculate the reward.
To be noticed, KES requires a learning record to initialize the learner's original knowledge level.

It's should be noted that neither of these two simulators is perfect. Each of them has its own limitation, i.e., the rule-based knowledge evolution in KSS possibly differs with the real-world and KES has the problem in describing the relations of knowledge structure. As these two simulators are complementary, the proposed approach should outperform others in these diverse environments at the same time to be proved robust. During simulation, the learners mastering all target items at the beginning will be skipped.

\begin{table}[t]
	\small
	\centering
	\caption{Characteristics of the comparison methods.}\vspace{-0.4cm}
	\label{tab:components}
	\begin{tabular}{|c|c|c|}
		\hline
		& Knowledge Level & Knowledge  Structure \\
		\hline
		KNN & $\times$ & $\times$ \\
		GRU4REC & $\times$ & $\times$ \\
		\hline
		MCS-10 & $\checkmark$ & $\times$ \\
		MCS-50 & $\checkmark$ & $\times$ \\
		DQN & $\checkmark$ & $\times$ \\
		CSEAL-NCN & $\checkmark$ & $\times$ \\		
		\hline
		CN-Random & $\times$ & $\checkmark$ \\
		\hline
		Cog & $\checkmark$ & $\checkmark$ \\
		CSEAL & $\checkmark$ & $\checkmark$ \\
		\hline
	\end{tabular}
\end{table}

\vspace{-0.5cm}
\subsection{Experimental Setup}

\subsubsection{Data Partition and Preprocessing.} The mentioned-above two simulators are applied as the environment for evaluating and on-line training. Due to their different characteristics, different data partition and preprocessing for them are applied.

\textbf{KSS.} As a rule-based simulator, KSS does not require any extra data to initialize it. Inspired by previous works ~\cite{chen2018recommendation,tang2018reinforcement,piech2015deep}, we employ KSS to generate the off-line dataset, i.e., \emph{dataOff}, which is then used to train  baseline methods or Knowledge Tracing model in the agents. The dataset \emph{dataOff} has 4,000 records with a max-length of 50. 

Without loss of generality, we randomly divide the dataset \emph{dataOff} into training, validation, and testing sets by the proportion of 80/10/10. The prerequisite graph in KSS contains 10 nodes and 12 links. The learning targets are randomly selected from the nodes.


\textbf{KES.} As a data-driven simulator, KES requires a certain amount of data for training DKT model so that we introduce the learning records mentioned in Section ~\ref{sec:DS} as \emph{dataSim}. Furthermore, we randomly divide \emph{dataSim} into two parts: \emph{dataOff} and \emph{dataRec} by the proportion 50/50. The details of three datasets are listed as follows:

(1) \emph{dataSim} is utilized to train the DKT model in the environment;

(2) \emph{dataOff} is used to train baseline methods and Knowledge Tracing model in the agents, which is the same as in KSS; 

(3) \emph{dataRec} is to retrieve the initialization records and learning targets. For each session record in \emph{dataRec}, the first 60\% of the data is applied to initialize the original learner's knowledge level in DKT. The middle 20\% is masked and the lasting 20\% is reserved as the learning target of this session.

We then divide each of those above datasets (i.e., \emph{dataDKT}, \emph{dataSimDKT} and \emph{dataRec}) for training, validation, and testing by the proportion of 80/10/10.

\vspace{-0.2cm}
\subsubsection{Framework Setting.}
Due to the different number of learning items in two simulators where 10 in KSS (a small number learning items scenario) and 835 in KES (a large number learning items scenario), different settings are set as follows:

\begin{itemize}[leftmargin=*]
	\item \textbf{KSS}: The embedding dimension used in DKT is 15, hidden dimension of LSTM is 20 while the dimension of output layer is the same as the number of learning items, i.e., 10. The dimensions of two layers in value-policy network are 128 and 32 respectively. 
	\item \textbf{KES}: The embedding dimension used in DKT is 600, hidden dimension of LSTM is 900 and the dimension of output layer is the same as the number of learning items, i.e., 835. The dimensions of two layers in value-policy network are 1,024 and 512 respectively. The DKT embedded in environment and agent shares the same value of dimension parameters but diverse training data.
\end{itemize}

\vspace{-0.3cm}
\subsubsection{Training Details.} 
We initialize parameters in all networks with \textit{Xavier} initialization ~\cite{glorot2010understanding}, which is designed to keep the scale of gradients roughly the same in all layers. The initialization fills the weights with random values in the range of $[-c, c]$ where c=$\sqrt{\frac{3}{n_{in} + n_{out}}}$. $n_{in}$ is the number of neurons feeding into weights, and $n_{out}$ is the number of neurons the result is fed to. We set mini-batches as 16. We also use dropout ~\cite{srivastava2014dropout} with the probability 0.2 for DKT embedding and each output in value-policy network and 0.5 for LSTM output to prevent overfitting and gradient clipping ~\cite{pascanu2013difficulty} to avoid the gradient explosion problem. Some of our codes are available in https://github.com/bigdata-ustc.

%

\begin{table}[t]
	\small
	\centering
	\caption{Overall results of $E_\mathcal{P}$.}\vspace{-0.4cm}
	\label{tab:results}
	\begin{tabular}{c c c c}
		\hline
		& KSS & KES \\
		\hline
		KNN & 0.000700  & 0.257919 \\
		GRU4Rec & 0.007727  & 0.201219  \\
		\hline
		MC-10 & 0.110577  & 0.002236  \\
		MC-50 & 0.108636  & -0.005103  \\
		DQN & 0.100610  & 0.002688  \\
		CSEAL-NCN & 0.222363  & 0.003354 \\
		\hline
		CN-Random & 0.272784  & 0.138526 \\
		\hline
		Cog & 0.164128  & 0.166560 \\
		CSEAL & \textbf{0.346883}  & \textbf{0.405823} \\
		\hline
	\end{tabular}
\end{table}

\vspace{-0.3cm}
\subsection{Baseline Approaches}
In order to demonstrate the effectiveness and robustness of our framework, we compare it with following methods.

\begin{itemize}[leftmargin=*]
	\item \textbf{KNN}: KNN ~\cite{cover1967nearest} finds a predefined number of learners nearest to the new learner by comparing the cosine distance of their learning paths, and decides what to learn next for the new learner.
	\item \textbf{GRU4Rec}: GRU4Rec is a classical session-based recommendation model ~\cite{DBLP:journals/corr/HidasiKBT15}. The input of the model is the sequence of the session while the output is the probability distribution of learning items which appear in the next step.
	\item \textbf{MCS}: Monte Carlo Search (MCS) ~\cite{metropolis1949monte} combined with KT is a searching method where KT predicts the promotion of each search path as the index for ranking.
	\item \textbf{DQN}: Some works ~\cite{chen2018recommendation, tang2018reinforcement} have proposed to leverage reinforcement learning for this problem, but these works require abundant human domain knowledge to design the transition matrix in MDP and an exact initial state, which is not practical. Thus KT model and Deep Q Leaning replace the states in MDP and simple q-learning separately.
	\item \textbf{CN-Random}: The recommended item is randomly picked from the candidate items selected by CN, and this baseline is treated as a simple knowledge structure based approach.
	\item \textbf{Cog}: The recommended item is weighted-randomly picked from the candidate items selected by CN,
	where the weight of the item is inversely proportional to the mastery measured by KT model.
	\item \textbf{CSEAL-NCN}: It is similar with our proposed model but without the Cognitive Navigation system.
\end{itemize} 

For better illustration, we summarize the characteristics of these models in Table ~\ref{tab:components}. All deep learning involved models are implemented by MXNet and trained on a Linux server with four 2.0GHz Intel Xeon E5-2620 CPUs and a Tesla K20m GPU.

\begin{figure} [t]
	\footnotesize
	\begin{center}
		\includegraphics[width=0.45\textwidth]{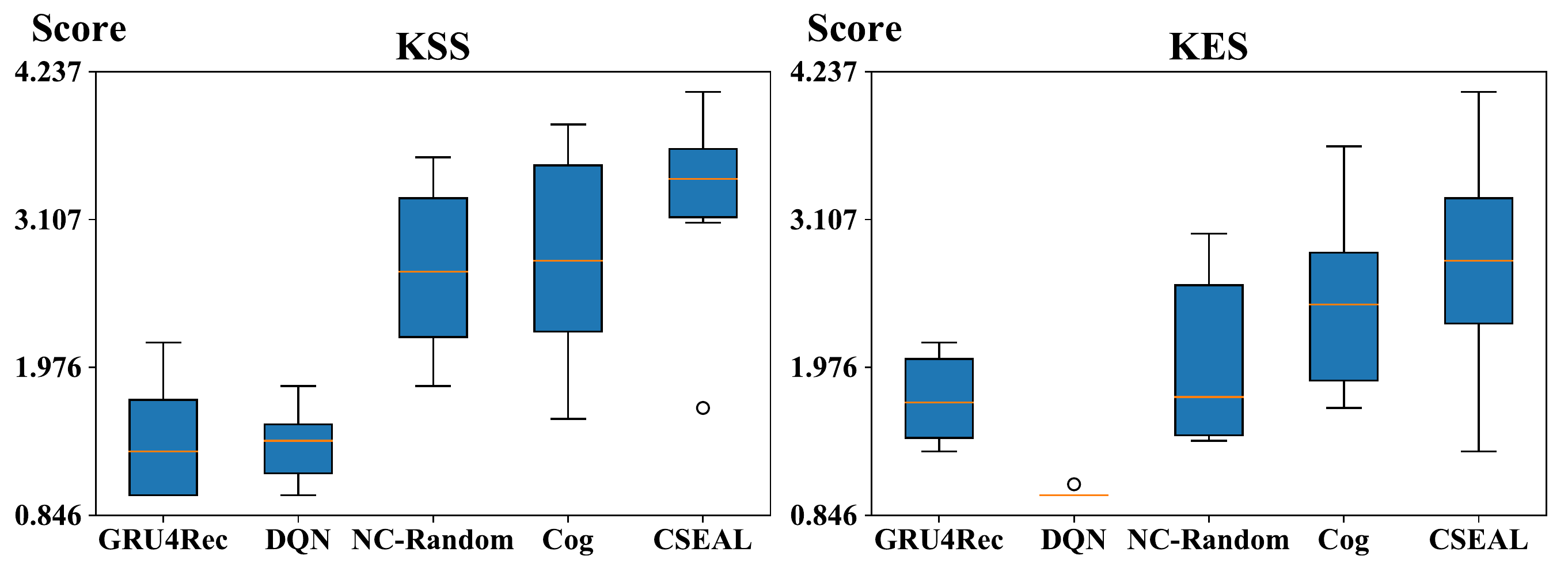}
	\end{center}\vspace{-0.45cm}
	\caption{Overall results of human study.}
	\label{fig:human}\vspace{-0.1cm}
\end{figure}

\vspace{-0.3cm}
\subsection{Evaluation Metrics}
Learning path recommendation focuses on the learning effectiveness rather than the selection of the learner (i.e., being practiced by the learner) or the correction of one exercise, which is essentially different from the general recommendation problem (e.g., merchandise recommendation, movie recommendation) ~\cite{zhu2018multi}. Because of this issue, classical metrics like precision, recall and NDCG can not be applied. However, the quantification of learning effect is not very clear and still a challenge ~\cite{salehi2014personalized}. Previous works use either the logicality of sequence ~\cite{zhu2018multi} or the promotion of knowledge level ~\cite{salehi2014personalized, chen2018recommendation} (i.e., $E_\mathcal{P}$) as the quantitative metric to evaluate the learning path. For more comprehensive results, we use both logicality and promotion as metrics for evaluation. We validate the performance of models particularly based on the promotion $E_\mathcal{P}$ given by simulators and the logicality evaluated by human experts. 

\vspace{-0.3cm}
\subsection{Experimental Results}
\subsubsection{Promotion Comparison.}
The length of recommended learning paths is set consistently to be 20 according to the median of exercises in one session estimated in Table ~\ref{tab:data}. Extensive experiments on the length of path will be conducted in Section ~\ref{sec:Length}. 

Table ~\ref{tab:results} shows the average $E_\mathcal{P}$ according to Equation (1) of all models. According to the results, obviously CSEAL performs best. Specifically, by modeling the knowledge levels, it beats CN-Random, GRU4Rec and KNN. By applying Cognitive Navigation on the knowledge structure, it achieves the better performance than DQN, MC-10, MC-50 and CSEAL-NCN. By well combining the knowledge level and the knowledge structure, it beats Cog. Then, in KSS, the methods with knowledge structure have better performance than the ones without knowledge structure which indicates that the knowledge structure contributes a lot to the effectiveness of recommended learning paths. Last but not least, in KES, GRU4Rec without explicitly modeling the knowledge level and the knowledge structure outperforms other methods except CSEAL. These observations infer that exploiting cognitive structure with the comprehensive consideration of knowledge level and knowledge structure for adaptive learning is necessary but challenging.

\begin{figure} [t]
	\footnotesize
	\begin{center}
		\includegraphics[width=0.43\textwidth]{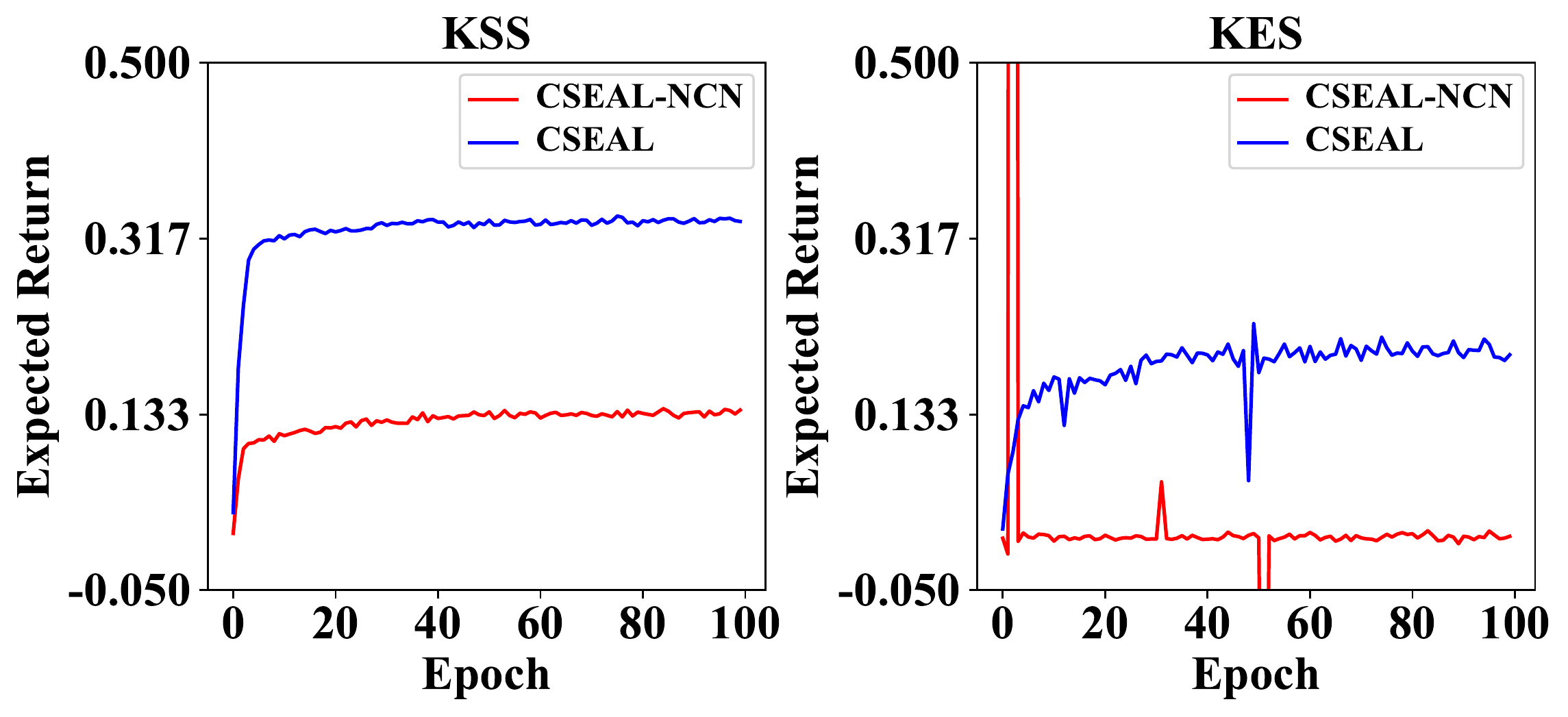}
	\end{center}\vspace{-0.5cm}
	\caption{Expected return in different learning epochs with or without Cognitive Navigation.}
	\label{fig:KGR}\vspace{-0.1cm}
\end{figure}

\vspace{-0.3cm}
\subsubsection{Experts Comparison.} \label{sec:EC}
Inspired by previous works ~\cite{zhu2018multi}, we invite six experts in education area who are familiar with learning path scheduling to evaluate the results of various methods based on their own logicality. Experts rate every learning path with a score from 1 to 5 and a higher score represents the higher logicality. Experts are asked to rate the 100 selected cases. Every case contains the historical learning record, the learning target and the recommended learning path. Four typical baselines, GRU4Rec, DQN, CN-Random and Cog, differing in whether having knowledge level or knowledge structure are selected as the comparison methods. 

As can be seen in Figure ~\ref{fig:human}, CSEAL outperforms all baselines. In other words, recommendations from CSEAL are in the most accordance with logicality of knowledge structure. Besides, it is observed that the methods with the Cognitive Navigation achieve a higher scores in experts evaluating, which indicates that the Cognitive Navigation helps to maintain the logicality of learning paths. Furthermore, two interesting phenomena draw our attention: (1) the scores of different experts for the same case sometimes are quite different; (2) compared with the former result of $E_\mathcal{P}$, we notice some models with higher score of logicality may not achieve better promotion. From these observations, the exact definition of logicality is not easy to be expressed and captured but varies in different people and it is additionally not equal to the promotion.

%


\vspace{-0.3cm}
\subsubsection{Impact of Knowledge Structure.} \label{sec:IKS}
Figure ~\ref{fig:KGR} presents the expected return obtained in each learning epoch of CSEAL and the variant CSEAL-NCN which does not have Cognitive Navigation. Notably, CSEAL is able to utilize the knowledge structure to obtain the optimal policy in a stable manner with the help of Cognitive Navigation. We can see CSEAL-NCN obtains a better policy in KSS than in KES because of the smaller searching space (i.e, smaller graph) in KSS. These issues indicate the knowledge structure can help reduce the searching space in reinforcement learning.

\begin{figure} [t]
	\footnotesize
	\begin{center}
		\includegraphics[width=0.45\textwidth]{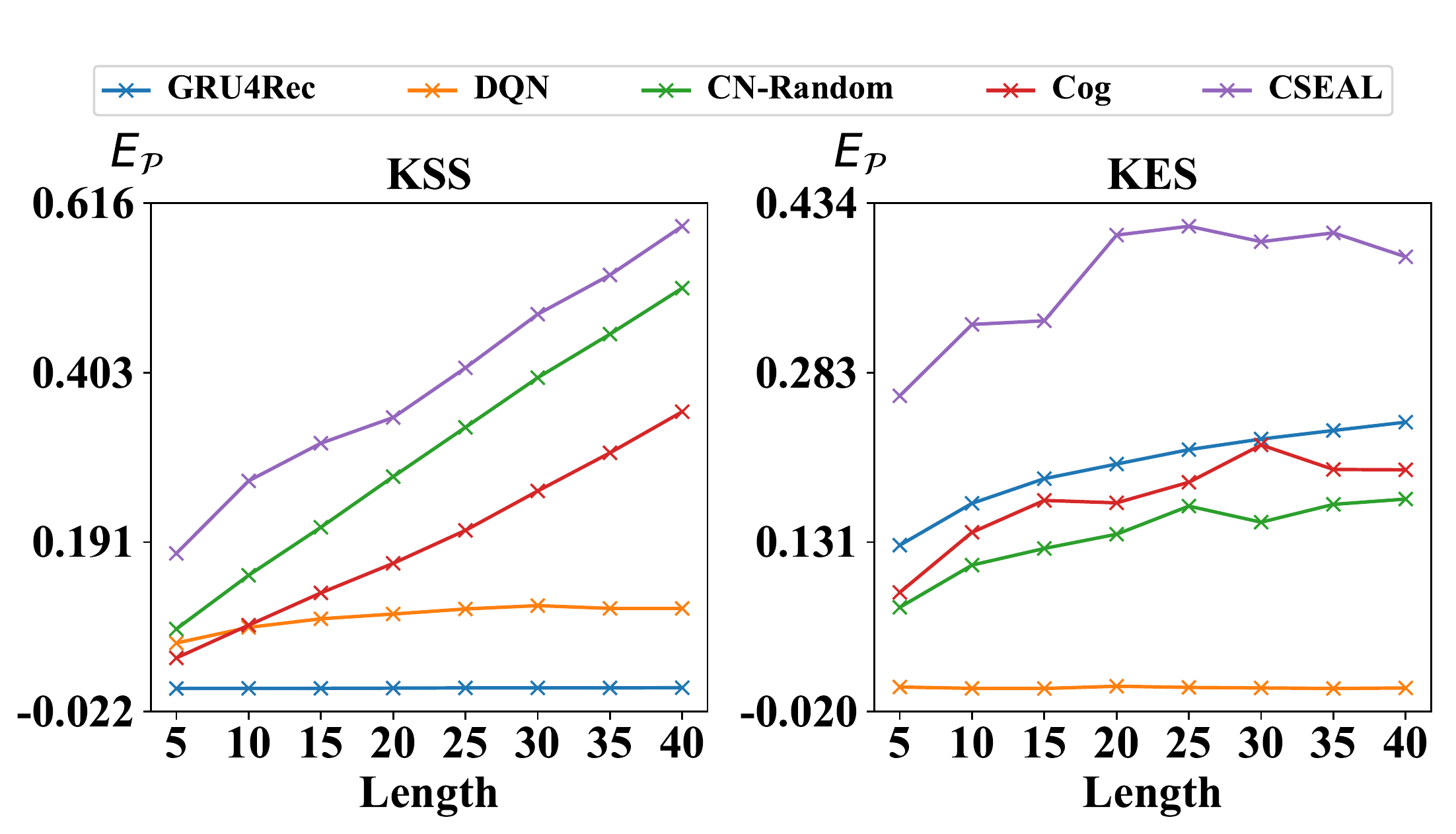}
	\end{center}\vspace{-0.5cm}
	\caption{Influence of learning path length.}
	\label{fig:Length}
\end{figure}

\begin{figure*} [ht]
	\footnotesize
	\begin{center}
		\includegraphics[width=0.85\textwidth]{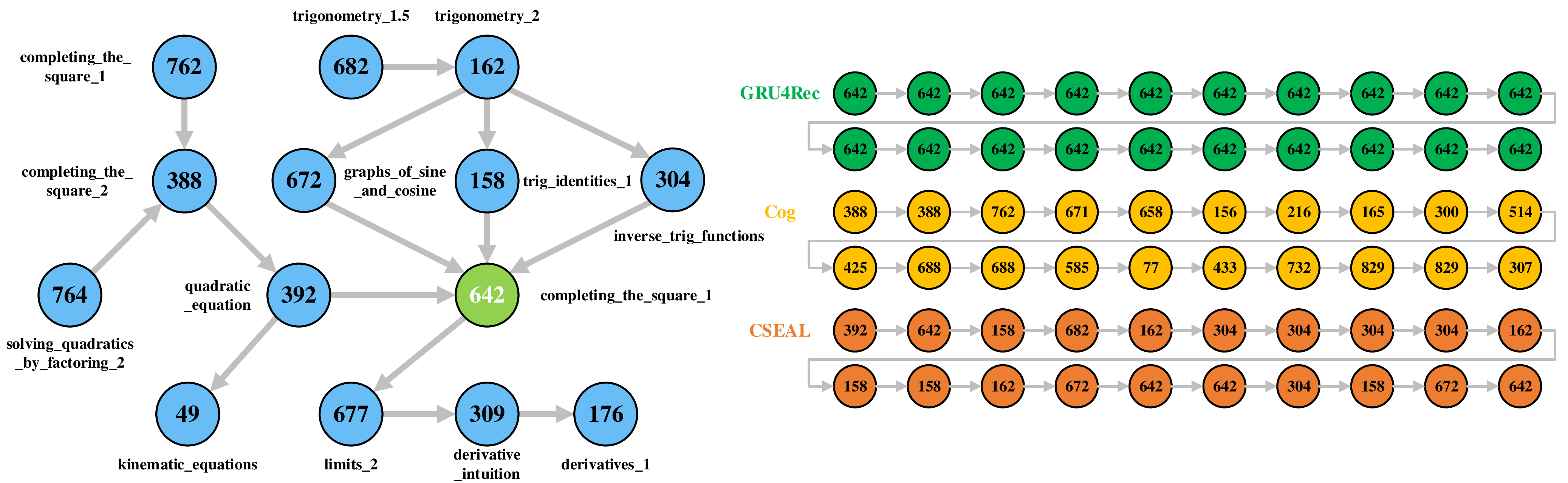}
	\end{center}\vspace{-0.45cm}
	\caption{Visualization of different recommended learning paths for the learning item 642, i.e., \textit{completing\_the\_square\_1}.}
	\label{fig:CaseStudy}\vspace{-0.6cm}
\end{figure*}


\vspace{-0.3cm}
\subsubsection{Performance with Different Length.} \label{sec:Length}
As shown in Section ~\ref{sec:DS}, the median of the length of a session is 21, from which we suspect that the most suitable learning length in KES should be near to it. We select the same methods in Section ~\ref{sec:EC} to verify our suspicion, and an extensive experiment is also performed on KSS. Figure ~\ref{fig:Length} shows the results. We can see that, in KSS, $E_\mathcal{P}$ grows with the length increasing because KSS is made by rules which have no limitation on length. While the effectiveness can hardly be promoted for all methods after 20 in KES, which verifies our suspicion. 

\vspace{-0.3cm}
\subsubsection{Case Study}
We give an example with visualization in Figure ~\ref{fig:CaseStudy}. In the example, a latest learning record of a learner who wants to learn the learning item 642, \textit{completing\_the\_square\_1}, is \{(642, 0), (642, 0), (642, 0), (642, 0), (642, 0)\} which illustrates that the learner possibly has some trouble in learning the target. To help this learner, different methods give different learning paths. For better understanding, we draw the subgraph containing three-hop neighbors of the target in Figure ~\ref{fig:CaseStudy}. GRU4Rec recommends a path where the learner continuously directly learns the target which he or she already seems to be stuck with. Cog recommends a path, where most learning items are far away from the target. Our method, CSEAL, encourages the learner firstly to review the prerequisite learning items and then go back to learn the target with reviewing the prerequisites. This visualization hints that CSEAL provides a more efficient and logical learning path for the learner to master the learning target.

\vspace{-0.3cm}
\section{Conclusions}
In this paper, we proposed a novel recommendation framework for adaptive learning, named CSEAL.
Specifically, based on the historical learning records, learning target and prerequisite graph, we firstly used a Knowledge Tracing model to retrieve the knowledge level of each learner.
Then, we applied a Cognitive Navigation system to maintain the knowledge structure of learning items.
Finally, we designed the Actor-Critic Recommender to dynamically provide learning items during a learning cycle.
The experimental comparisons with seven baseline methods on two Simulators (i.e., KSS and KES) with diverse scenarios and human experts clearly demonstrated both the effectiveness and robustness of our framework. As a general framework, each step of CSEAL may be further improved in the future. 


\textbf{Acknowledgements.} This research was partially supported by grants from the National Key Research and Development Program of China (No. 2018YFC0832101), the National Natural Science Foundation of China (Grants No. 61672483, U1605251), and the Science Foundation of Ministry of Education of China \& China Mobile (No. MCM20170507). Qi Liu gratefully acknowledges the support of the Young Elite Scientist Sponsorship Program of CAST and the Youth Innovation Promotion Association of CAS (No. 2014299).

\bibliographystyle{abbrv}
\bibliography{CSEAL}

\end{document}